\documentclass{iau}

\usepackage{amsmath}
\usepackage{graphicx}
\usepackage{multirow}

\begin{document}

\lefttitle{Creaner, O. et al., 2024}
\righttitle{Proceedings of the International Astronomical Union.}

\jnlPage{1}{7}
\jnlDoiYr{2024}
\doival{10.1017/xxxxx}

\aopheadtitle{Proceedings IAU Symposium}
\editors{eds.}

\title{The Exoplanet Citizen Science Pipeline: Human Factors and Machine Learning}

\author{Ois\'{i}n Creaner, Anna Preis, Cormac Ryan \& Nika Gorchakova}
\affiliation{Dublin City University}

\begin{abstract}
We present the progress of work to streamline and simplify the process of exoplanet observation by citizen scientists. International collaborations such as ExoClock and Exoplanet Watch enable citizen scientists to use small telescopes to carry out transit observations. These studies provide essential supports for space missions such as JWST and ARIEL. Contributions include maintenance or recovery of ephemerides, follow up confirmation and transit time variations. Ongoing observation programs benefit from a large pool of observers, with a wide variety of experience levels. Our projects work closely with these communities to streamline their observation
pipelines and enable wider participation. Two complementary approaches are taken:
Star Guide applies human-centric design and community consultation to identify
points of friction within existing systems and provide complementary online tools and
resources to reduce barriers to entry to the observing community.
Machine Learning is used to accelerate data processing and automate steps which are
currently manual, providing a streamlined tool for citizen science and a scalable
solution for large-scale archival research.
\end{abstract}


\maketitle

\section{Introduction}
\label{Introduction}
We work to remove barriers to entry for citizen scientists. Our work addresses both technical and human factors which impede participation. By removing these barriers, we permit a larger number of people to participate in the study of exoplanets.

Astronomy has long been a field where amateur observers can make a significant contribution.  Even in the era of the James Webb Space Telescope (JWST) and forthcoming Atmospheric Remote-sensing Infrared Exoplanet Large-survey (ARIEL) missions, amateur observers remain indispensable. Amateur astronomers can enhance exoplanet ephemerides \citep{noguer2024enhancing}, or visually inspect light curves with \textit{Planet Patrol}, one of many Zooniverse-hosted projects \citep{kostov2022planet}. By gathering large pools of observers, citizen science projects can bypass the bottleneck of limited time and availability of professional observatories \citep{peluso2023unistellar}. The forthcoming Vera C. Rubin provides a pipeline directly from the Rubin Science Platform to the Zooniverse to support scientists in successfully running crowd science projects, including those in the field of Exoplanets \cite{schwambenabling}. 

We collaborate formally with Exoplanet Watch \citep{zellem2020utilizing} and ExoClock \citep{kokori2022exoclock}, citizen science projects which enable members of the public to contribute observations to exoplanet science. These projects have identified  areas which pose a challenge, especially to inexperienced observers. Our ongoing work will elucidate areas of difficulty and work to address those challenges.  

Our team includes an academic, undergraduate and postgraduate students and a web and software development professional without a background in astronomy.  This diverse group has been essential in identifying areas for improvement which would not be obvious to a more traditional, purely academic group. For example, jargon terms familiar to a professional (e.g. ephemerides, limb darkening) may be unknown to citizen scientists. These terms can be off-putting and thus act as a barrier to entry.

\section{Current Pipeline}
\label{Pipeline}
\begin{figure}[ht]
    \centering
    \includegraphics[width=0.75\linewidth]{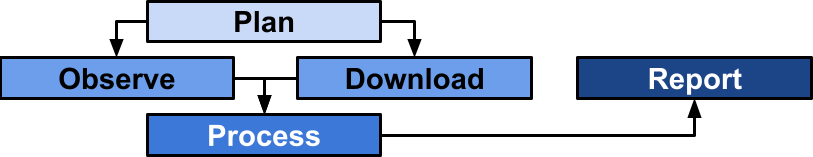}
    \caption{The workflow of a citizen scientist working on exoplanet transits. \textbf{Plan}: select a target. \textbf{Observe} a target directly or \textbf{Download} data from a previous observation. \textbf{Process} the raw images into a lightcurve. \textbf{Report} the results by uploading the lightcurve.}
    \label{fig:pipeline}
\end{figure}

The process by which citizen scientists observe exoplanets can be treated as a pipeline as illustrated in Figure \ref{fig:pipeline}. Observers must first plan their observation, selecting the star to observe determining when to make the observation. Observers must then gather data, either by carrying out the observation themselves or by use of remote facilities. They must then process that data at home, generating a light curve with software on their home computer.  Finally, they submit their results to a central repository, where they can be collated with results from across the collaboration. The details of this process vary by project, but the overall principles are consistent \citep{zellem2020utilizing, kokori2022exoclock, peluso2023unistellar}. 

\textbf{Plan}:
In the planning phase, the observer must determine a target to observe. Tapir \citep{Jensen2013Tapir} (also known as the Swathmore Transit Finder) and ExoWorldsSpies Transit Scheduler \citep{Tsiaras_Kokori_2017} are online tools for identifying observable transits. They both permit an observer to input parameters such as their location, equipment and time of day/year, and will return lists of transits which will be observable.

To mitigate against atmospheric effects which would otherwise dominate over the small effect of a transit, techniques such as differential photometry and temporal binning are used \citep{hartley2023optimized}. Careful selection of reference stars can mitigate against wavelength-dependent effects such as atmospheric reddening and techniques such as the Locus Algorithm \citep{creaner2022locus} can help identify optimum fields of view to observe.

\textbf{Observe}:
Observers typically use Charge-Coupled Device (CCD) or other digital cameras to capture images, and computer-controlled telescope mounts to keep the object in frame. Many citizen science projects rely on highly heterogeneous equipment to gather their data\citep{zellem2020utilizing, kokori2022exoclock}. The UniStellar project is an exception, relying on the Enhanced Vision Telescope (eVscope), potentially simplifying data processing \citep{peluso2023unistellar}.

\textbf{Download}:An alternative approach is for citizen scientists without access to a telescope (and/or clear, dark skies) to make use of remotely controlled telescopes such as MicroObservatory \citep{fowler2021observing}. This can broaden the pool of potential observers. Data gathered remotely is processed identically to data gathered locally.

\textbf{Process}:
Gathered data is typically stored in Flexible Image Transport System (\texttt{FITS}) files, a lossless data storage system designed for astronomy \citep{wells81fits}. The data is processed using a variety of tools, such as EXOTIC (EXOplanet Transit Interpretation Code) \citep{zellem2020utilizing} and HOPS (HOlomon Photometry Software, \citet{kokori2022exoclock}).  These tools typically require user input to identify the target and reference stars, and the selection of these is not always obvious. These software packages process the images into lightcurves.

\textbf{Report}:
These lightcurves can be visually inspected by the observer to approximate factors such as transit depth and signal-to-noise ratio (SNR). The underlying data, together with a report on the instrument and software used to process it, are uploaded to a remote server. There, they are combined with data from other observers to produce aggregate data with greater fidelity \citep{zellem2020utilizing, kokori2022exoclock, peluso2023unistellar}.

\section{Human Factors}
\label{Human}
\begin{figure}[ht]
    \centering
    \includegraphics[width=0.65\linewidth]{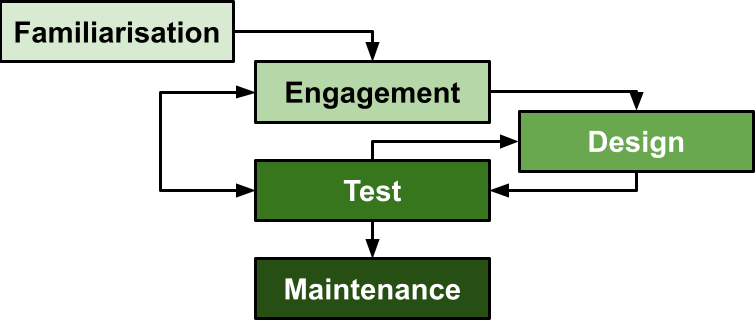}
    \caption{The workflow of the group in addressing human factors in exoplanet citizen science projects. First, we \textbf{familiarise} ourselves with the systems. Next, we \textbf{engage} with the community to identify their needs. We then \textbf{design} solutions to these problems in an iterative cycle which includes \textbf{testing} and continued engagement with observers. Finally, we intend to launch and \textbf{maintain} the solutions developed.}
    \label{fig:approach}
\end{figure}

It can be seen from the above pipeline that there is significant direct participation by the citizen scientist in the observation process. While participation is the strongest form of public engagement, it can prove daunting for inexperienced observers. The make-up of our team as outlined in Section \ref{Introduction} allows us to readily identify these barriers. Our work builds on industry expertise and the principles of design affordance to streamline the pipeline wherever possible. Our approach has five key phases as shown in Figure \ref{fig:approach}.

\textbf{Familiarisation}: Internal use of existing systems by members of the project team gives us a basis to understand the processes currently followed. This also ensures sufficient familiarity with the systems to interpret comment and feedback from the community.

\textbf{Community Engagement}: We Work with the existing user community to identify points of friction and areas with potential for improvement. This takes the form of both informal interviews and anonymous community surveys carried out using online tools. The results of these surveys are expected to be available in Q3 of 2024.

\textbf{Design Solutions}: Iterative design processes will allow new solutions to be developed with a rapid feedback cycle. This includes work on user guides \citep{Preis2024Guide}, website content \citep{Preis2024Website} and software  \citep{Creaner_Stellar_Matchmaker_2024}, which are currently in progress.

\textbf{Community Testing}: Throughout the development of solutions, we remain engaged with community members to ensure their needs are met. This includes the recruitment of alpha- and beta-testers for software and discussions at community meetings.

\textbf{Launch and Maintenance}: Due to  constrained timelines for this project, when we finalise and launch any content, it will be intended to function with limited maintenance. However, we will continue to monitor performance and seek feedback on the project while  practicable.

\section{Machine Learning}
\label{ML}
Machine Learning (ML) is an umbrella term for a category of computer algorithms and models which ``learn" from their input data \cite{el2015machine}, allowing computers to perform tasks such as data analysis and prediction with minimal explicit programming \cite{sharifani2023machine}. ML solutions offer improvements in areas we categorise as the \textbf{Quality} of outcomes and the system \textbf{Performance}.

\textbf{Quality}: An ideal ML model can be defined by the phrase ``Probably Approximately Correct” model, coined by \citet{valiant1984theory}. A model is said to be ``Approximately Correct" if its output is within an acceptable margin of error from the true result. A ``Probably Correct" model has a high probability of producing the correct result over a large sample of predictions \cite{hausslerprobably}. The definitions of ``acceptable margin" and ``high probability" are application-sensitive. For exoplanet science, because transits are rare, we value \textit{precision} $(\frac{True\; transits}{Transits\; predicted})$ over \textit{accuracy} $(\frac{Correct\; predicitons}{Sample\; size})$. One method to assess these quality factors is to divide input data into ``Training" and ``Validation" samples. Training data is used by the model to learn how to map inputs to outputs. The Validation data is used to assess the model outputs against known outputs by determining how \textit{probable} it is that the model is \textit{approximately} correct \cite{choi2020introduction}.

\textbf{Performance}: ML has significant potential to accelerate a variety of workflows in exoplanet observation \cite{jara2020transiting}. A key point in accelerating a process is to identify the bottlenecks in the process. In general, manual processes are slower than automated processes, and ML algorithms tend to outpace programmed solutions. Benchmarks such as CPU/GPU time per process, memory consumption and so on are used to assess performance \cite{takamoto2022pdebench}.

We have identified several areas of the exoplanet pipeline in which relatively little ML research has been focused, but which show potential for growth. Key among these is computer vision: using machine learning techniques to support extraction of light curves from data by minimising the effect of noise on the outcome. Another area of potential is observation planning: using archives and catalogues together with machine learning technology to plan efficient observing campaigns. Key to evaluating the success of these approaches will be the development of relevant metrics against which a ML solution can be evaluated.

\section*{Acknowledgements}

Star Guide (OC, AP, CR) is funded by the National Open Research Forum (NORF) Open Research Fund 2023, Strand II: Open Research Stimulus.

NG and OC are funded by Science Foundation Ireland through the SFI Centre for Research Training in Machine Learning (Grant No. 18/CRT/6183) Call 2023. This funding is provided through the Higher Education Authority (HEA).

\bibliographystyle{plainnat} 
\bibliography{Sample}
\end{document}